\documentclass[%
 reprint,
superscriptaddress,
 amsmath,amssymb,
 aps,
]{revtex4-2}
\usepackage{color}
\usepackage{gensymb}
\usepackage{graphicx}
\usepackage{dcolumn}
\usepackage{bm}
\usepackage{hyperref}
\usepackage{epstopdf}
\usepackage{float}
\usepackage{listings}
\usepackage{times,mathptm,bm}
\usepackage{xcolor}
\usepackage[export]{adjustbox}
\usepackage{upgreek}

\input ulem.sty\relax
\catcode`\@=11 %

\font\uwavefont=lasyb10 scaled 652

\def\uwave{%
  \bgroup
    \markoverwith{%
      \lower3.5\p@\hbox{\uwavefont\char58}%
    }%
  \ULon
}

\begin{document}

\title{Worm Blobs as Entangled Living Polymers:\\ From Topological Active Matter to Flexible Soft Robot Collectives}


\author{Antoine Deblais}
\affiliation{Van der Waals-Zeeman Institute, Institute of Physics, University of Amsterdam, 1098XH Amsterdam, The Netherlands.}
\author{K.~R.~Prathyusha}
\affiliation{School of Chemical and Biomolecular Engineering, Georgia Institute of Technology, Atlanta, GA 30332.}
\author{Rosa Sinaasappel}
\affiliation{Van der Waals-Zeeman Institute, Institute of Physics, University of Amsterdam, 1098XH Amsterdam, The Netherlands.}
\author{Harry Tuazon}
\affiliation{School of Chemical and Biomolecular Engineering, Georgia Institute of Technology, Atlanta, GA 30332.}
\author{Ishant Tiwari}
\affiliation{School of Chemical and Biomolecular Engineering, Georgia Institute of Technology, Atlanta, GA 30332.}
\author{Vishal~P.~Patil}
\affiliation{Department of Bioengineering, Stanford University, Stanford CA 94305 US.}
\author{M.~Saad Bhamla}
\affiliation{School of Chemical and Biomolecular Engineering, Georgia Institute of Technology, Atlanta, GA 30332.}
\date{\today}

\begin{abstract}
Recently, long and slender living worms have garnered significant interest because of their impressive ability to exhibit diverse emergent behaviors in highly entangled physical and topological conditions.
These worms can form an active viscoelastic, three-dimensional soft entity known as the ``blob'', which can behave like a solid, flow like a liquid, and even respond to external stimuli such as light to locomote or change shape.
To understand the behavior of the blob, it is crucial to consider the high degree of conformational entanglement that individual units can achieve because of their high aspect ratio and tunable activity.
This topologically active collective necessitates reevaluating established soft matter concepts in polymer physics to advance the development of active polymer-like materials. Our understanding of the complex emergent dynamics of the worm blob promises to catalyze further research into the behavior of entangled active polymers and guide the design of synthetic topological active matter and bioinspired tangling soft robot collectives.
\end{abstract}

\pacs{Valid PACS appear here}
\maketitle

An essential ingredient that enables individuals to achieve more is their ability to interact with one another, ranging from transient interactions to physically entangled systems. The nature and timescale of these interactions can differ widely among both living and artificial individuals, resulting in a diverse range of complex behaviors. In general, when displaying self-repulsive interactions, individuals remain unconnected, resulting in fluid-like behavior, such as in human crowds,\cite{Bain2019} flocking birds,\cite{BirdflockEmlen1952} or schooling fish. When interactions between active individuals become attractive, large cohesive structures can form that exhibit new mechanical responses, such as elastic behavior in rafts of ants\cite{Mlot2011} or aggregates of living cells.\cite{Mehdiabadi2006}

\begin{figure}
	\centering
\includegraphics[width=0.9\linewidth]
 {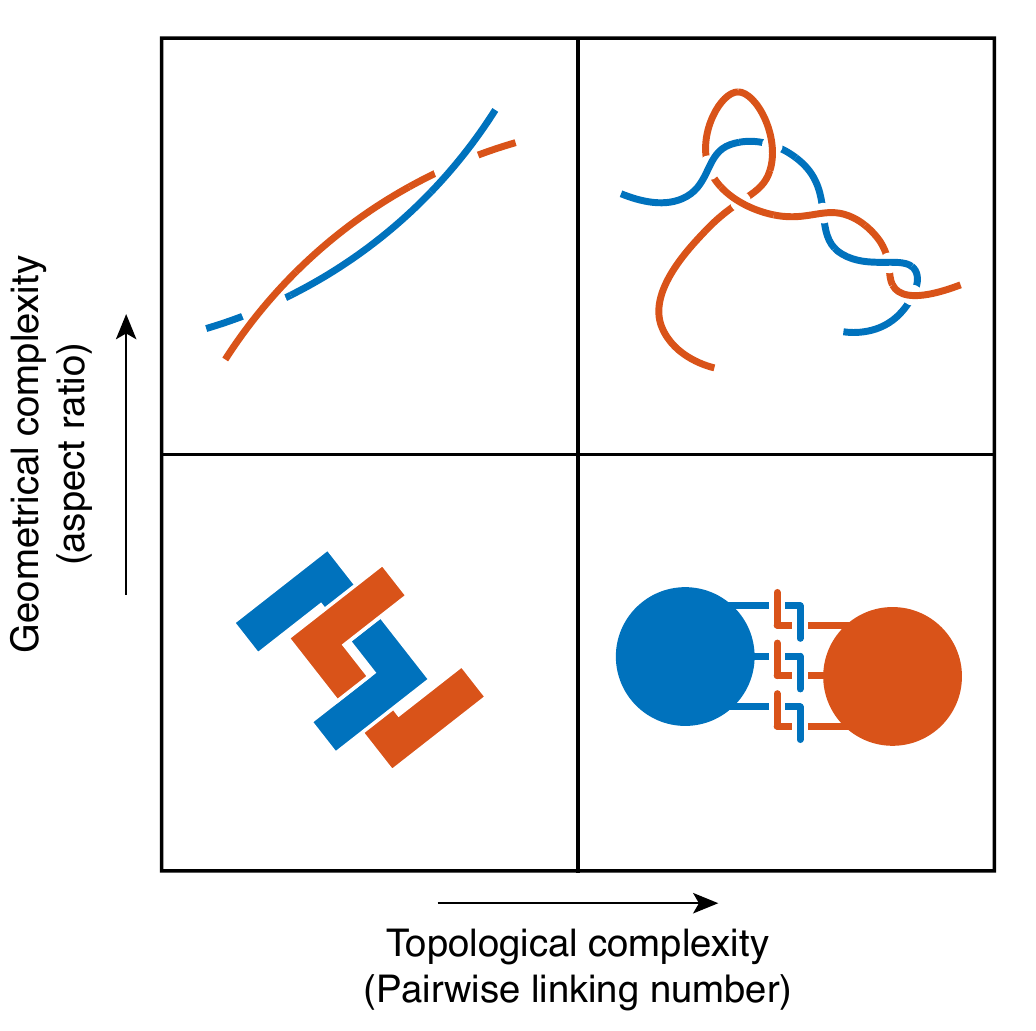}
	\caption{{\bf Classification of tangled matter.} The components of entangled matter are classified according to their aspect ratio, and the complexity of the topological structures they can form. Rigid U-shaped particles\cite{Gravish2012} (bottom left) and the long rigid sticks (top left) which make up birds' nests\cite{Weinernestmechanics2020} typically only form single braids before breaking. Particles with hooks (bottom right), which capture ant-like entanglement,\cite{Mlot2011} can form multiple links despite their low aspect ratio. Due to their flexibility and large aspect ratio, worm-like filaments (top right) are capable both of forming topologically complex structures and exhibiting long-range interactions.
	}
	\label{fig:entangle-cohesion}
\end{figure}

It is important to differentiate between transient attractive interactions and physical entanglement. Entanglement in soft matter traditionally refers to braiding and linking between filament-like objects, from stiff, jointed rods\cite{Mlot2011} to flexible polymers and filaments.\cite{deGennes1971,deGennes1976} Although these constituent elements interact through repulsive contact forces, entanglement gives rise to a configurational trap, thereby generating the effective attractive forces which give entangled matter its remarkable stability. Such entangled systems can be further classified by the topological and geometrical properties of their components (Fig.~\ref{fig:entangle-cohesion}). Aspect ratio is a measure of geometrical complexity,\cite{Weinernestmechanics2020} and captures the extent to which the contact interactions are long-range. Topological complexity\cite{shankar2022topological}, as measured by quantities such as the linking number,\cite{patiltuazon2023} quantifies the complexity of the braid-like structures which can be formed. In \textit{passive} soft matter, a classical and distinguished example of a highly topologically entangled system is a polymer solution: a liquid composed of flexible microscopic constituents with a long-aspect ratio that can form strong physical entanglements. These physical entanglements give rise to the unique properties found in polymer solutions.~\cite{deGennes1971,deGennes1976,Ingremeau2013} Owing to their geometric and topological complexity, active worm collectives occupy a unique region in the phase space of tangled matter (Fig.~\ref{fig:entangle-cohesion}), and thus have the potential to exhibit new, functional forms of collective behavior.

\begin{figure*}
\centering
  \includegraphics[width=0.8\linewidth]{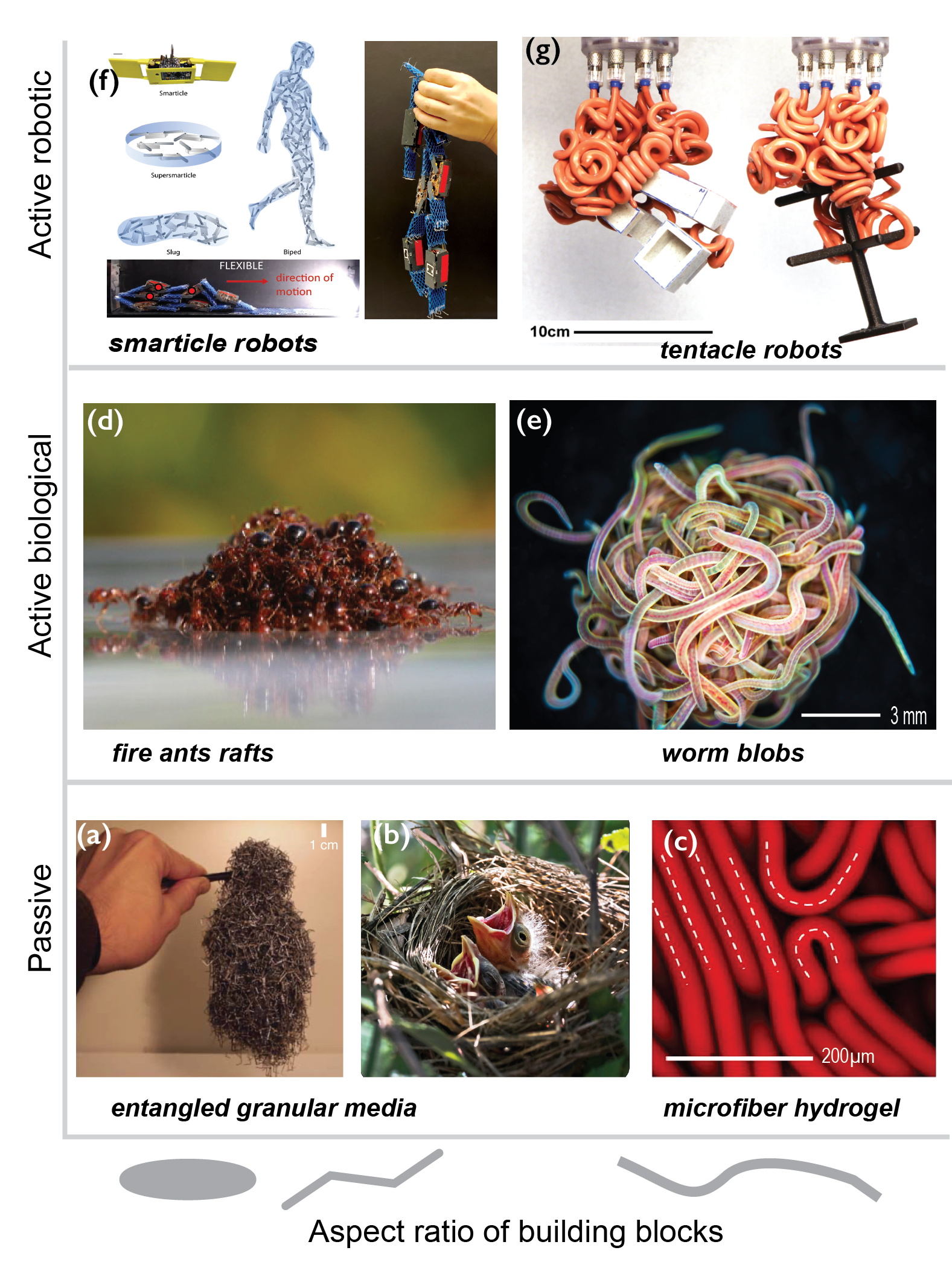}
  \caption{{\bf Physically and topologically entangled matter.} From passive systems to active and soft robotics: in both the passive and active worlds, an increase in the particle aspect ratio leads to entanglement, and new macroscopic mechanical behaviors can emerge.
  (a) When increasing the aspect ratio of the particle, new physical properties can emerge as observed for a collection of staples [extracted with permission from Gravish \textit{et al.}]\cite{Gravish2012}, where a (partial) degree of entanglement leads to a solid-like aggregate. (b) A bird's nest constructed from branches and twigs, a natural engineering marvel, is an example of an entangled cohesive granular structure, having remarkable properties such as plasticity despite its elastic elements and softness even though its filaments are hard.~\cite{Weinernestmechanics2020} (c)
  Irreversible formation of topological entanglements upon shear leading to a dramatic change in the rheological properties of the very-high-aspect ratio flexible fiber suspension.\cite{Perazzo2017}
  (d) In the living world, fire ants benefit from a partial degree of entanglement enabling them to build complex structures such as rafts\cite{Mlot2011} and survive a flood. (e) The worm blob exhibits a high degree of topological entanglement which is at the core of a plethora of emergent behaviors that may inspire future direction in soft robotic physics.\cite{patiltuazon2023} (f) Robophysical model of the worm blobs displaying the  emergent locomotion in the collective entangled state.\cite{Ozkan2021} (g) Similar to how a jellyfish collects a fish, a soft robotic gripper uses the collection of active thin tentacles or filaments to entangle and ensnare objects.\cite{Beckerstentaclerobot2022}}
  \label{fig:PanelEntangledActiveMatter}
\end{figure*}

At the nano- and micro-scale, biology offers numerous instances of \textit{active} polymer structures, spanning from actin filaments and microtubules, which constitute the main components of the cytoskeleton in living cells,\cite{Koenderink2006,Kirchenbuechler2014} to flagella found in sperm, algae, bacteria, and various other planktonic microorganisms. These active systems reap the benefits of entanglement, whereby the topological entanglement of actin filaments contributes to the cytoskeleton's distinct properties. Comprehending the non-equilibrium statistical mechanics of active systems poses a significant challenge both theoretically and experimentally. Although recent advancements have been made in theory,\cite{ghosh2014dynamics,Winkler2017,Bianco2018,Martin2018,Locatelli2021} there are limited synthetic experimental systems available for active filaments, which are often restricted to a small number of basic entities such as driven colloidal particles attached together,\cite{yan2016reconfiguring,biswas2017linking,Stuij2019} or self-propelled (ro)bots.\cite{Li2021,Zheng2021} These systems can be challenging to manipulate or access in large quantities.

Recent research has overcome these experimental challenges by utilizing active entities that rely on living biological worms: the California blackworm \textit{(Lumbriculus variegatus})\cite{Ozkan2021,Nguyen2021,patiltuazon2023} and the sludge worms \textit{Tubifex tubifex}.\cite{Deblais2020a,Deblais2020b} These studies demonstrate that the motion and dynamics of these worms can be analyzed and their activity can be easily controlled with simple methods such as temperature manipulation\cite{Ozkan2021,Nguyen2021} or the addition of alcohol.\cite{Deblais2020a} This makes living worms excellent candidates to investigate the behavior of active polymers in various situations.\cite{Deblais2020a,Deblais2020b,Ozkan2021,Nguyen2021} When dispersed in large quantities of water, these worms can spontaneously aggregate \cite{Deblais2020b} into highly entangled states, forming large assemblies or ``blobs'' that closely resemble a melt of regular polymers.\cite{Deblais2020a,Ozkan2021} Once entangled, the worms collaborate and exhibit vibrant, unexpected behaviors following P.~W.~Anderson's precept ``More is different''.\cite{Anderson1972}

Here, we present our perspective on this new type of biological living polymer particle both as an individual entity and as a large assembly, known as the blob. We make the case that these living worms provide an outstanding experimental platform for investigating the physics of active polymer-like particles and call for a re-examination of classical polymer concepts. This perspective paper showcases the richness and emergent behaviors resulting from the combination of activity, long aspect ratio, and entanglement of the living polymer-like worms (as shown in Fig.~\ref{fig:PanelEntangledActiveMatter}). It highlights the enormous potential of this living system to achieve complex tasks autonomously, such as shape-shifting, a dream that has long been the subject of science fiction (as seen in the 1958 classic, The Blob). The worm blob opens up new possibilities and should inspire various communities, ranging from soft matter physicists to soft roboticists.

\section*{Topologically and physically entangled worms: The blob}

\subsection*{The \textit{Lumbriculus variegatus} and \textit{Tubifex tubifex} worms: Two remarkable model organisms}
In the biological kingdom, both blackworms and sludge worms belong to the phylum Annelida, which is a diverse group of segmented worms that inhabit a wide range of environments (marine, freshwater, and terrestrial habitats) and play an important ecological role in decomposition and nutrient recycling via bioturbation. These worms can be repeatedly cut into more than a dozen pieces, and each segment will regenerate into a complete worm by architomy fission.\cite{Martinez2021} Annelids have a lengthy history of study, dating back to the 18$^{th}$ century, with a focus on their remarkable regenerative abilities. For a rich and detailed history, including past biological pioneers and the incredible animal itself up to 2021, readers can consult Rota\cite{Rota2022} and Martinez \textit{et al.}\cite{Martinez2021}. Additionally, we especially want to highlight Charles Drewes' passionate and inspiring contribution in proposing blackworms as a resilient and accessible organism for teaching and for research.\cite{Drewes1990,Drewes1999,Lesiuk1999}

Both oligochaete worms are commonly used as live food for fish, prawns, or frogs and can easily be found in pet shops. Both worms naturally inhabit freshwater and reside in the sediment of lakes and rivers, although \textit{T.~tubifex} can also be found in sewer lines. A panic event was reported in North Carolina when colonies of sludge worms were observed (see the press article and video in \citet{Wired2009}). Both  have been the subject of intensive biological studies due to their ability to thrive in harsh environments, and they are commonly used as indicators of polluted environments.\cite{Marian1984,Khangarot1991,Hurley2017}
Due to their unusual slime and wiggling behavior in low-oxygen conditions, \textit{T.~tubifex} were dubbed ``the sewer creatures''.

As individuals, these worms seemingly look like a conventional polymer with a long chain comprised of repeated segments that display wiggling motion and allow the worm to be self-propelled and crawl on a surface. Both worms are approximately 0.3-0.5~mm thick and 10-50~mm long (depending on their habitat, age, and nutrition), giving them an aspect ratio of $\sim $100. The thermal random motion of the worms is negligible compared to their active motion, so they constitute a simple model system for active polymers and they can be cultured, harvested, and analyzed with cheap and simple tools.

The crawling motion, dynamics, and conformations of these worms have been analyzed in detail.\cite{Deblais2020b,Ozkan2021,Heeremans2022} Sitting on a surface, \textit{T.~tubifex} motion is a random walk with an effective diffusion constant that increases with the temperature of the surrounding environment, as it has been confirmed by extracting the mean square displacement from which a long-time diffusion coefficient has been retrieved.\cite{Deblais2020b} In contrast, \textit{L.~vareigatus} displays more ballistic movement, given that their surrounding temperature is not excessively high.\cite{Nguyen2021}

Both worms live in the benthic regions of bodies of water and prefer cool and dark environments. They are ectothermic and mainly rely on their surrounding environment for thermoregulation. Additionally, they have photoreceptors along their tails to detect potential predators.\cite{drewes1989hindsight} Therefore, they exhibit both negative phototaxis and negative thermotaxis and have been observed to locomote away from these regions.\cite{Ozkan2021}  

\subsection*{Physics of active filaments}

\begin{figure*}
\centering
  \includegraphics[width=0.8\linewidth]{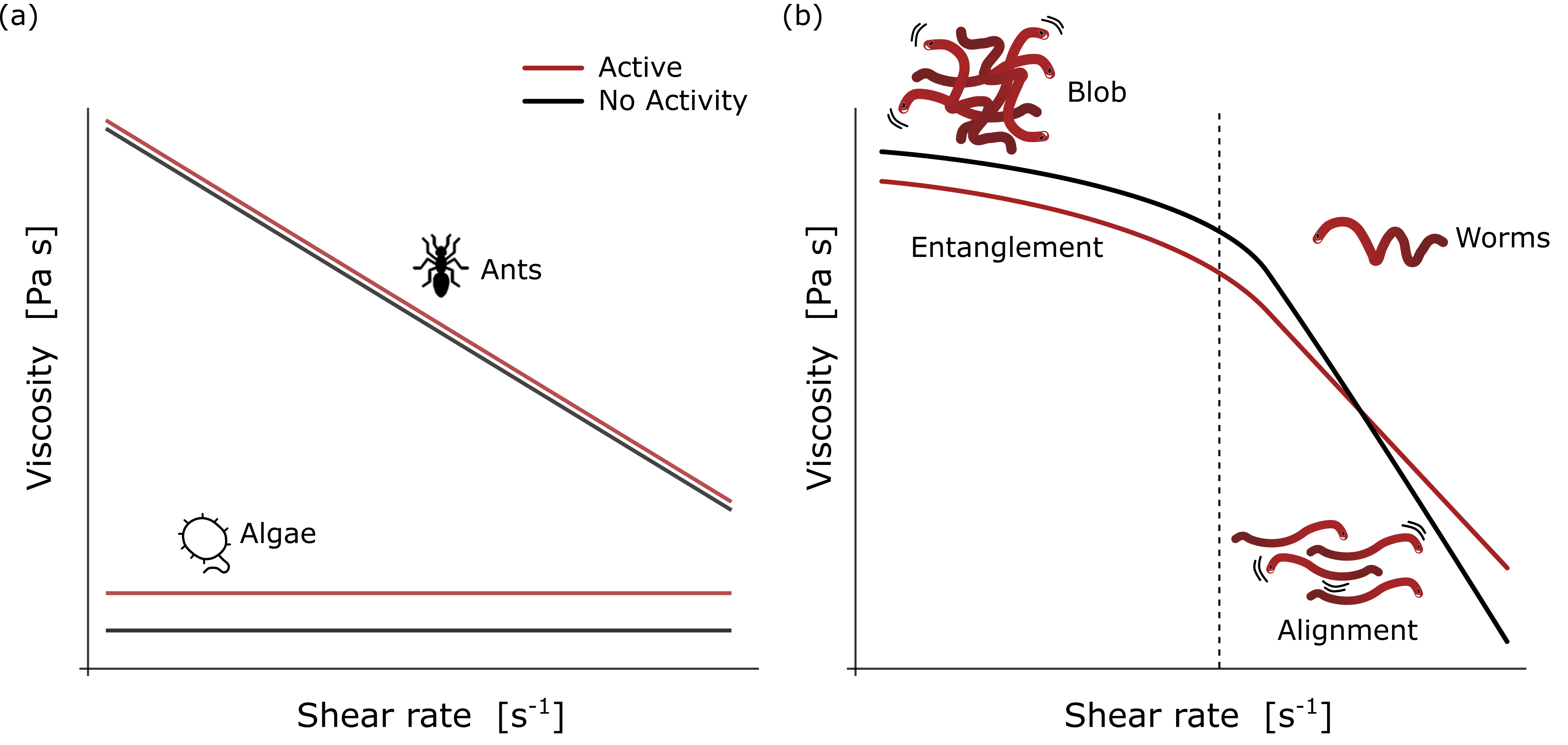}
  \caption{{\bf Rheological fingerprint of the blob:} (a) Schematic representation of the flow curves (shear viscosity as a function of shear rate) for ants\cite{Tennenbaum2016} and algae\cite{Gachelin2013} for two different activity states [active (red) Vs non-active (black)]. (b) By analogy to conventional polymers, the rheology of a blob exhibits two distinct regions at low and high deformation rates with a strong dependence on activity. Adapted with permission from Deblais \textit{et al.}.\cite{Deblais2020a} }
  \label{fig:Rheologyblob}
\end{figure*}

The worm is an elongated and slender living organism capable of moving its body by internal mechanisms (peristaltic motion),\cite{alexander2003principles,kudrolliPNAS2019,quillin1999peristaltic} making it a suitable example of an active polymer.\cite{Deblais2020a,Deblais2020b,Nguyen2021} Numerous other examples of active filaments exist in the living world, and most studies have focused on motor-driven cytoskeletal filaments,\cite{ganguly2012cytoplasmic,fjnedelec-97,ysumino-12,llgof-02,farkas-02,tsanchez-12} DNA/RNA during the transcription process,\cite{guthold1999dna-rna} cilia\cite{gilpin2020multiscale} or flagella,\cite{mayfield1977rapid} sperm,\cite{friedrich2010sperm} rolling viruses,~\cite{sakai2018unique} parasites,~\cite{klug2017motility} bacteria~\cite{berg2004coli} and snakes.~\cite{rieser2021functional-snake} A common feature of these systems is the interplay between the activity, flexibility, and conformational degrees of freedom that gives rise to a wide range of structural and dynamical properties at the individual polymer level\cite{riseleholder-15,gayathri-PRL-12} and the collective.\cite{winkler-gompper-review-activepolymerJCP-20,winkler_active_2017,farkas-02,prathyusha2018PRERastko,manna2019emergent,nasirimarekani2022active}

Identifying the underlying propulsive mechanisms or processes that are responsible for the activity of these filaments is crucial in developing computational\cite{patra2022collective,prathyusha2018PRERastko,prathyusha2022transverse,leiladirectionreversal2023,gayathri-PRL-12} and theoretical models \cite{ziebert2021influenza,harvey2013continuum,liverpool2001viscoelasticity,Martin2018} for active polymers. Since the worms exhibit a wiggling motion and self-propelling mechanism, the closest realization might be a tangentially propelling polar active filament model. In this model, each polymer monomer propels with a constant force parallel to its local tangent. The model successfully predicts the single active polymer dynamics\cite{hiang-14a,hiang-14b,riseleholder-15} and collective pattern formation exhibited by cytoskeletal filaments and bacteria.\cite{prathyusha2018PRERastko,leiladirectionreversal2023,duman2018collective} 

Although hydrodynamic interactions between the filaments are known to influence the motion and coordinated movement of active polymers,\cite{hiang-14b,gayathri-PRL-12} experimental observations of worms suggest that such interactions are negligible.\cite{Deblais2020b} This indicates a dry active polar filament model is enough to represent a worm motion. 

In the collective, the worms form a blob by tangling their slender bodies with each other. Such entanglement gives intriguing rheological properties when shear is applied to the worm blob (see next section).\cite{Deblais2020a} The temperature-dependent entangled network within the worm blob actively contributes to its non-Newtonian fluid behavior, enabling it to flow over long timescales while preserving its solid shape during short timescales.~\cite{Ozkan2021,Deblais2020a} Although visco-elastic and rheological properties of active polymers have been a focus of various theoretical and computational studies,\cite{liverpool2001viscoelasticity} the worm blobs warrant further research owing to their resemblance to a polymer melt with complex activity patterns. 

Current models of active polymers primarily focus on 2D motion, imitating the flexible living organisms that propel or glide on surfaces or interfaces. However, the formation of topologically tangled networks within 3D blobs necessitates the adoption of 3D polymer models to truly grasp their morphological dynamics. Existing theoretical work has introduced minimal models consisting of chains of active particles with specified active force directions, aiming to capture the behavior of these microscopic organisms. Although limited by its two-dimensionality and simplified self-propulsion, this model serves as a foundation for future research.

By expanding these models to encompass three dimensions and incorporating hydrodynamics, we can further investigate the roles of individual sensing and environmental interactions in cooperative searching behavior. Inspired by these tiny organisms, upcoming theoretical and computational studies promise to deepen our understanding of active polymers, transcending the boundaries of conventional passive polymer physics. Such advancements hold the potential for unlocking new applications in materials science and biophysics while shedding light on the rich physics underlying these tangled active matter systems.

\subsection*{The worms blob}

For biological purposes, these living worms may spontaneously aggregate to minimize their exposure to an excess of oxygen dissolved in water depending on their metabolic requirements.\cite{Walker1970} They form compact 3D-aggregates or highly entangled blobs, a process similar to polymer phase separation, and for which we recently measured the kinetics of aggregation.\cite{Deblais2020b} The growth occurs by the coalescence of smaller aggregates into larger ones through strong interactions - entanglement - of the individual worms. Interestingly, the coalescence between the blobs is possible because the worm blobs themselves are capable of moving. Similar to the individual motion of a worm, the blob exhibits a random walk. Surprisingly, and in stark contrast with regular polymer solutions or colloids subject to Brownian motion, measurements reveal that the diffusion coefficient of the blob is independent of its size. This is possible because the worms inside the entangled blob are effectively immobilized, and only the worms on the outer surface of a blob contribute forces.~\cite{patiltuazon2023}

\section*{Emergent properties of a worm blob}

\subsection*{Rheological \& mechanical properties}

\textbf{Effect of activity.} Inanimate materials are typically classified based on their mechanical response to external forces. Solid materials are known for exhibiting little change over time or when subjected to external forces, while liquids are strongly influenced by external stresses and can easily flow under the force of gravity. However, viscoelastic materials are much more common and exhibit complex behavior that falls between these two extremes.

Remarkable advances have been made in the study of \textit{active} material systems, such as dilute suspensions of algae\cite{Gachelin2013} and aggregates of fire ants,\cite{Tennenbaum2016} which have provided compelling evidence that activity can dramatically enhance the shear viscosity of these systems (Fig.~\ref{fig:Rheologyblob}(a)) compared to their inactive counterparts. These findings have important implications for our understanding of the mechanical properties of materials and may lead to new approaches for the design of high-performance materials. In particular, the shear rheology tests performed on fire ant aggregates \cite{Tennenbaum2016} have revealed a degree of shear-thinning, where the cycles of assembly and disassembly of bonds (legs) between the active constituents lead to stress relaxation. Although the viscosity of the system varies significantly over several decades, no significant effect is observed once the fire ants are rendered inactive (Fig.~\ref{fig:Rheologyblob}(a)). This can be attributed to the fact that in active systems like this, viscosity is primarily determined by the friction between the legs of the individuals, which highlights the importance of considering the collective behavior of active materials when designing new materials with desired mechanical properties.

\begin{figure*}
 \centering
 \includegraphics[width=1\linewidth]{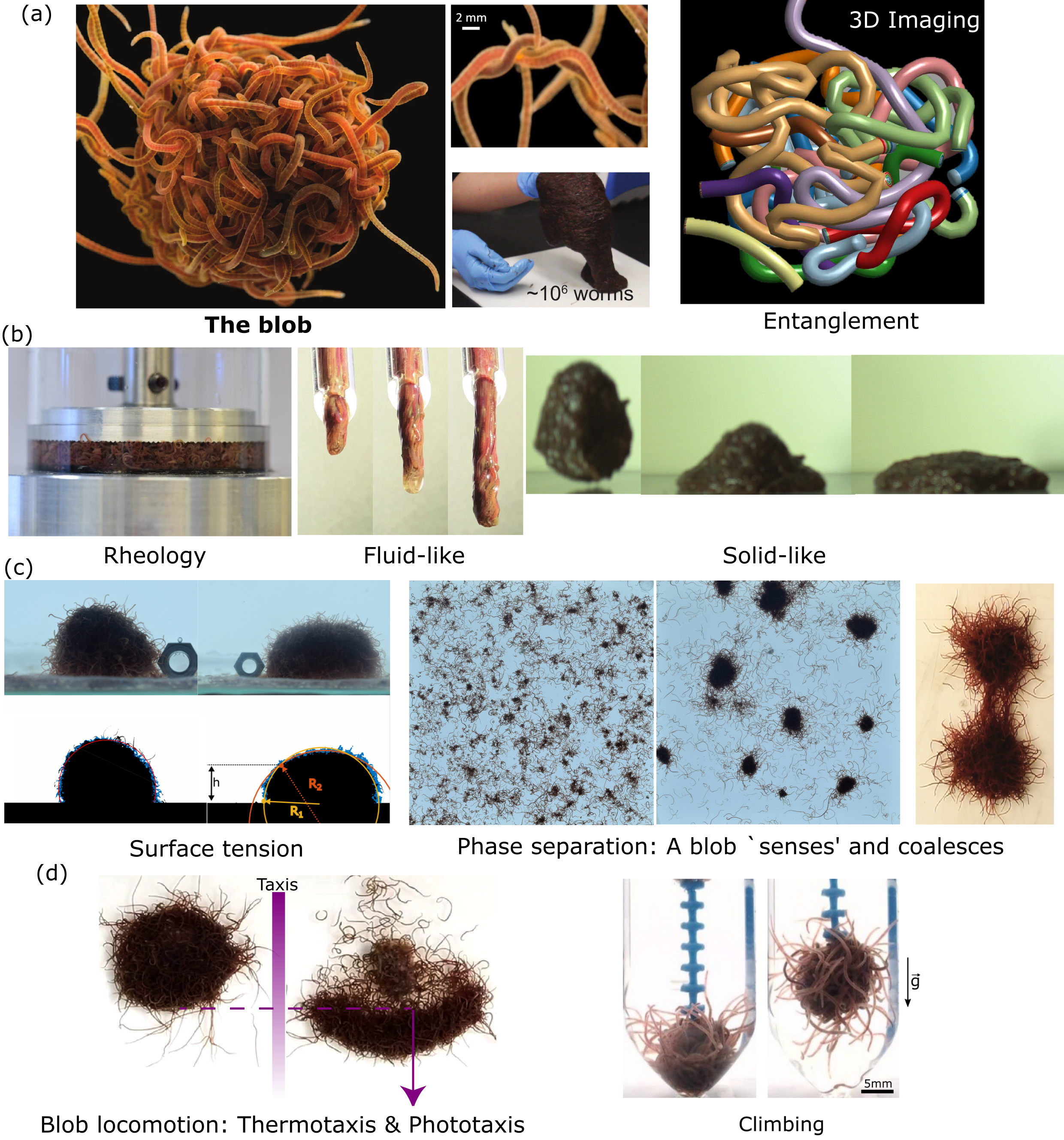}
 \caption{{\bf Emergent properties of a blob at a glance.} (a) A blob is made of highly entangled worms as shown in the picture and reveal from 3D-Imaging reconstruction. (b) Mechanical/rheological properties: The blob behaves as a liquid and can bounce as a solid. (c) In analogy to wetting phenomena and polymer solutions, living worms form a blob through phase separation by entanglement (reversible) and coalescence with an effective surface tension. (d) The blob is even able to respond to external stimuli such as light, temperature, or oxygen to generate emergent locomotion.}
 \label{fig:Blobataglance}
\end{figure*}

The results of our rheology measurements on living worms (Fig.~\ref{fig:Rheologyblob}(b)) provide compelling evidence of the unique and fascinating behavior of active polymer systems. By comparing our living worms to conventional polymer solutions, we were able to demonstrate a striking attenuation of the shear-thinning behavior due to the activity of the worms. This effect is in stark contrast to the behavior of conventional polymer solutions, where increasing flow rate leads to a decrease in flow resistance. Furthermore, we found that changes in activity levels result in unique rheological behavior that has not been observed in any other living polymer-like system. In particular, increasing activity levels led to a significant decrease in zero shear viscosity, a phenomenon that is not seen in conventional polymer solutions. These results demonstrate the rich and complex behavior of active materials and provide important insights into the underlying physics governing their rheological properties.

\textbf{Effect of oxygen concentration}. Another interesting aspect of the worm blob is its changing stiffness and shape as a function of the dissolved oxygen (DO) present in the water. 
As aerobically respiring organisms, the oxygen consumption rate is another key parameter that worm blobs must balance along with their level of entanglement. It was previously observed that a 1g blob of blackworms ($\sim$150 individuals) can consume enough oxygen in a 20 mL volume of water to create an anoxic ($<$ 1 mg/L) environment for themselves in around 30 minutes.~\cite{tuazon2022oxygenation} As a result, the individual worms wave their tails around away from the blob to supplement their oxygen supply. This in turn lowers the structural rigidity of the blob for low DO concentration.~\cite{tuazon2022oxygenation}

Subsequently, our recent work measured the change in the tensile strength of the blob as a function of DO concentration.~\cite{savoie2023amorphous}
Due to the worms' waving their tail in low oxygen environments, we found that the blob could exert a tensile force of almost 3 times larger in magnitude when the DO concentration was high ($>$ 8 mg/L) compared to anoxic conditions ($<$ 1 mg/L). Furthermore, we also observed that blobs of worms in ample oxygen behave like rigid objects which can topple over, while their counterparts in anoxic environments behave more like a viscoelastic gel. Therefore, blackworm blobs resemble active entangled matter whose stiffness is tunable by exogenous stimuli.

\subsection*{Emergent locomotion}

The collective locomotion of worm blobs displays emergent behaviors that go beyond binary gait differentiation. Worms within the blob can intertwine to form braided chains and pull together.\cite{Ozkan2021}
Experiments measuring the pulling force exerted by individual worms showed that a small number of worms could generate sufficient traction force to move the collective away from negative stimulus. Additionally, locomotion in small blobs can emerge through the differential activities of individual worms at the front (puller worms generating forward thrust) and rear (wiggler worms reducing friction). These results were further validated through robophysical models using multiple three-link phototactic smarticles.\cite{Ozkan2021}

Building upon the previous study, we created a model using the physics of active, semi-flexible polymers and filaments.\cite{Nguyen2021} The model simulates worms as self-propelled Brownian polymers, focusing on the parameter space of aspect ratio, bending rigidity, activity, and temperature. Simulated single worms display persistent directed motion at low temperatures, while multiple simulated worms can aggregate into a blob, held together by attractive forces. The study finds that the blob can collectively navigate along a temperature gradient, provided that the tangential force and attachment strength are balanced. We observe a tradeoff between worm velocity and blob cohesiveness, with an ``optimal'' regime identified for effective collective locomotion in the form of a phase diagram.\cite{Nguyen2021}

Perhaps even more interestingly, the worm blob as a whole can break symmetry and locomote across a substrate in response to chemicals, light, temperature, and fluidic gradients (see Fig.~\ref{fig:Blobataglance}(d)). The modes of locomotion of the blob can be of different natures depending on the external stimulus. When the worm blob is placed in a temperature gradient, a crawling motion is observed. In this scenario, worms facing the colder temperature extend outside the blob and pull on it, while the worms experiencing the hotter temperatures on the other side wiggle their bodies to reduce friction underneath the blob, facilitating its transport.~\cite{Ozkan2021,Nguyen2021}
      
Thus, the blobs can crawl, form floating structures~\cite{Tuazon2023Buoy} and even shape-shift to undergo complex topological transformations from pancakes to spheres. Although Fig.~\ref{fig:Blobataglance} highlights that many complex behaviors are possible, we believe that many other complex behaviors are yet to be discovered that may further depend on the blob size, the activity, the type of external stimulus or the substrate properties.

\subsection*{Reversible tangle topology}

\begin{figure*}
	\centering
	\includegraphics[width=\linewidth,trim={2cm 0 2cm 0},clip]{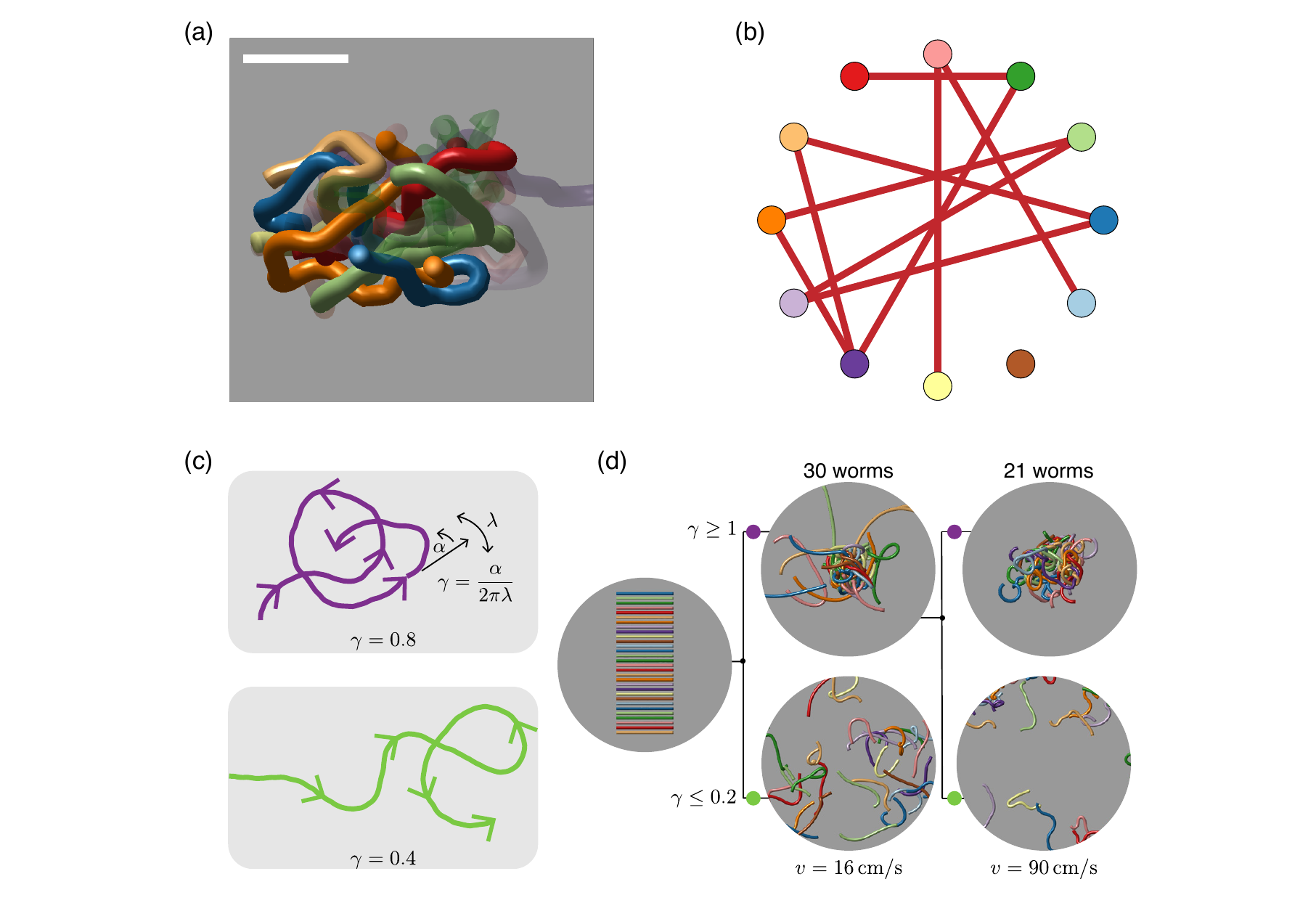}
	\caption{{\bf Topological structure and topological dynamics of worm blobs.}
	(a)~Ultrasound reconstruction of a living worm blob enables key topological interactions to be identified. Scale bar 5mm.
	(b)~Tangle graph captures the topological state of the worm blob.
	(c)~Mean-field picture of worm tangling focuses on worm head trajectories (purple and green curves). The chirality number $\gamma = \alpha/2\pi\lambda$ determines whether a worm head trajectory winds around obstacles more (top row) or less (bottom row). Trajectories with more winding (large $\gamma$) lead to tangled states.
	(d)~Simulations of worms with different values of $\gamma$ confirm that $\gamma$ controls the emergent topological states, even at different speeds. Worms with large $\gamma$ produce tangled states (top row), whereas worms with small $\gamma$ produce untangled states (bottom row). Each worm has length $40\,$mm and radius $0.5\,$mm.
	}
	\label{fig:tangle_topology}
\end{figure*}

Due to their ability to form and escape from tangled states, worm blobs represent a model system in which to explore questions of topological dynamics of tangled filaments. A fundamental question that must be addressed before analyzing dynamics concerns classifying and quantifying the topological state of a tangle. Ultrasound imaging methods, which enable the reconstruction of the internal, 3D structure of a worm blob (Fig.~\ref{fig:tangle_topology}a), provide insight into this problem. These datasets reveal that worm blobs are strongly interacting systems; each worm touches almost every other worm. Specific key interactions can further be selected on topological grounds. For example, pairs of worms that touch each other do not interact as strongly as pairs of worms that both touch and intertwine. Using the topological quantity known as the linking number, this observation can be made more precise. The topological state of the worm blob may then be approximated by a tangle graph, where an edge is drawn between two nodes if the corresponding worms both touch and are sufficiently intertwined (Fig.~\ref{fig:tangle_topology}b).

The topological dynamics of worms moving in complex tangles can be approximated and understood using a mean-field theory (Fig.~\ref{fig:tangle_topology}c), in which the head of each worm moves in 2D. Intuitively, trajectories that wind more (Fig.~\ref{fig:tangle_topology}c, top row) lead to tangled states, whereas trajectories that avoid one-way winding (Fig.~\ref{fig:tangle_topology}c, bottom row) lead to untangled states. The amount of winding is captured by the chirality number, $\gamma$, a dimensionless number that plays a key role in the topological dynamics of worms. The chirality number relates the rate at which the worm head turns, $\alpha$, with the rate at which the worm switches from turning left (right) to turning right (left), $\lambda$. Trajectories with large $\gamma$ (Fig.~\ref{fig:tangle_topology}c, top row) wind more than those with small $\gamma$ (Fig.~\ref{fig:tangle_topology}c, bottom row). The chirality number, therefore, provides a mechanism for controlling the emergent topology of a worm collective. Multi-filament numerical simulations in 3D confirm that the chirality of worm trajectories determines the emergent topological state formed (Fig.~\ref{fig:tangle_topology}d). Worms with large $\gamma$ tangle, whereas worms with small $\gamma$ untangle, thus establishing a connection between dynamics and topology. The topological tools developed thus far for quantifying tangle topology and predicting tangling dynamics promise to provide a platform for understanding more complex layers of worm blob functionality. Furthermore, these tools could aid in the design of robotic worm blob analogs.

\section*{Discussion and perspectives}

\subsection*{Living world-inspired polymer concepts: A brief history}

\begin{figure}
	\centering
	\includegraphics[width=\linewidth]{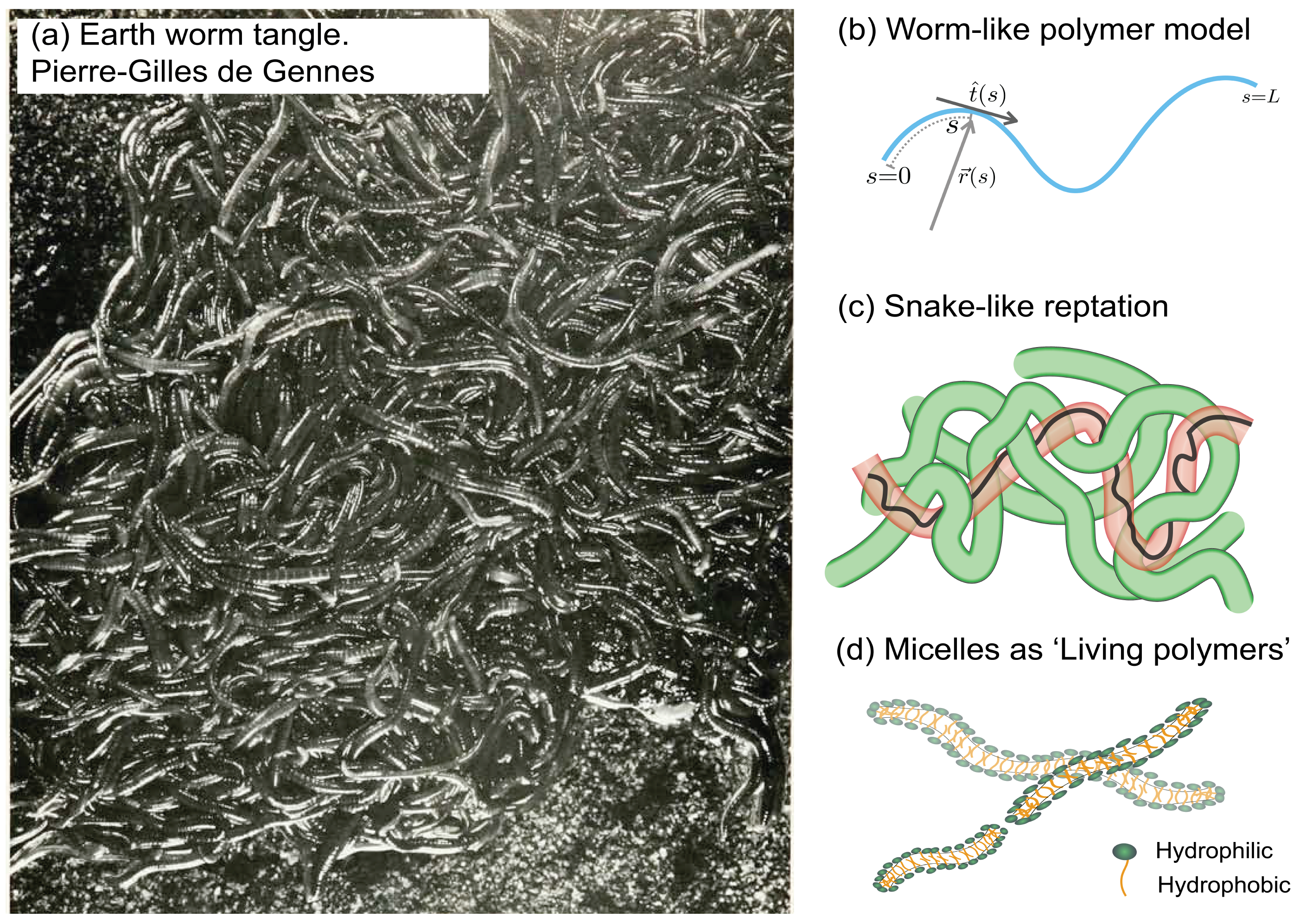}
	\caption{{\bf Classical concepts in polymer physics inspired by living systems.} (a) Pierre-Gilles de Gennes  used the analogy of a ``tangle of long earthworms'' to describe the motion of long polymer molecules in a melt, drawing inspiration from the movement of these worms. Adapted from Physics Today (1983).\cite{worm-degennes-83} (b) The ``worm-like model'' represents a long and semi-flexible polymer chain of length $L$, parameterized by the local tangent vector $\hat {t}(s)$. This concept was developed by Kratky and Porod (1949).~\cite{kratky-porod1949,hermans1952statistics} (c) Pierre-Gilles de Gennes (1971), Masao Doi, and Sam Edwards (1978) introduced the concept of ``reptation'', which describes the movement of an entangled polymer chain through a ``tube'' formed by adjacent chains.~\cite{deGennes1971,doi1978dynamics,doi1978dynamics-2,doi1978dynamics-3} (d)  Michael Cates (1987)  developed the theoretical  "living polymer"  model for worm-like surfactant assemblies where the chains undergo thermally induced  scission-recombination kinetics.\cite{cates1987reptation,cates1988dynamics}}
	\label{fig:living-worldpolymer}
\end{figure}

In this perspective, we directly study worms as active polymers, acknowledging the long, rich history of biological systems like worms and snakes inspiring classical polymer physics ideas. Although previous work has been inspired by these organisms, they primarily served as conceptual aids without involving actual experiments or modeling. Early advancements in polymer physics were significantly influenced by these biological systems, resulting in essential theoretical models and concepts (Fig.~\ref{fig:living-worldpolymer}). Nobel Prize-winning physicist Pierre-Gilles de Gennes famously drew an analogy between a tangle of earthworms and the behavior of polymer melts, consisting of long, intertwined molecular chains.~\cite{worm-degennes-83} The ``reptation theory'' describes the unique motion of a polymer through these entangled chains as the major relaxation mechanism, initially proposed by de Gennes in 1971~\cite{deGennes1971,degennes1979scalingconcept} and later expanded to the tube model by Masao Doi and Sam Edwards.~\cite{doi1978dynamics,doi1978dynamics-2,doi1978dynamics-3,doi1988book} This theory suggests that a polymer chain in the melt exhibits ``snake-like'' motion within a virtual tube formed by surrounding chains, restricting its free movement, much like snakes slithering among one another. Recently a few groups have been looking into the effect of internal activity on the reptation of polymers, either directly,~\cite{tejedor_dynamics_2020,tejedor2023molecular} or by studying simulations of active polymers in porous environments.~\cite{mokhtari_dynamics_2019,kurzthaler_geometric_2021} Experimental systems have mostly focused on bacteria or point particles\cite{bhattacharjee_bacterial_2019,Bechinger2016} using these worms would be a first truly polymer-like experimental system to check the predictions that have come from the theoretical and computational studies mentioned in this paper.

In 1949, Kratky and Porod,\cite{kratky-porod1949} described threadlike molecules as chains composed of elongated, cylindrical segments, leading to the development of a continuum ``worm-like'' polymer model for semi-flexible polymers.~\cite{hermans1952statistics,daniels_1952, saito1967statistical} The model accounts for stiffness via the inclusion of bending elasticity and found applications in investigating a wide range of polymers, including natural biopolymers such as proteins and DNA and synthetic polymers.  In the late 20$^{th}$ century, researchers discovered that surfactant molecules could form elongated structures resembling long polymers.~\cite{candau1985light,imae1985formation} Unlike ordinary polymers, these worm-like chains exhibit thermally induced scission and recombination dynamics on experimental time scales. Called ``living polymers'' due to the reversible breaking of chains at equilibrium, these structures are inspired by growth and division processes observed in living organisms. In the 1980s, Michael Cates became the first to integrate models of entanglement with the reversible breaking dynamics of these ``living polymers'',~\cite{cates1987reptation,cates1988dynamics,turner1991linear} spurring further research in the field.~\cite{huang2006kinetics,lerouge1998shear,prathyusha2013shear,thakur2010shear} The study of active polymers today holds the potential to uncover new principles in non-equilibrium polymer physics and inspire the development of innovative technologies, paralleling how \textit{passive} polymer physics has been influenced by the living world.

\subsection*{Topologically entangled living polymer physics}

Our recent findings on semi-aquatic worm blobs have important implications for the field of soft matter physics, shedding light on the emerging area of ``topologically entangled living polymers.'' As a living material, the worm blob demonstrates the intricate coupling of bulk mechanics and morphological computation. Notably, these active, entangled living polymers, which can dynamically modulate their rheology and topology, share similarities with synthetic polymer solutions that exhibit tunable physical properties. This positions worm blobs as a promising experimental platform for investigating the physics of out-of-equilibrium polymers.

The worm blob presents a unique opportunity to revisit the principles of soft matter physics, such as entanglement, active reptation, rheological plasticity, and fluid-structure interactions, by incorporating activity into these concepts. Additionally, the exploration of reversible tangle topology and the development of topological tools for quantifying tangle dynamics can help advance our understanding of the worm blob's functionality. This approach has the potential to reshape the field of soft matter physics and provide valuable insights into the behavior of topologically entangled living polymers.

\subsection*{Towards soft robotic blobs}

We envision soft, slender, spaghetti-like filamentous robots that transition topologically from floppy individuals to cohesive, emergent, task-capable soft robotic ensembles. This concept is reminiscent of the science fiction depicted in ``The Blob'', and we anticipate that studying actual worm blobs will help turn this concept into reality. Investigating these worms could potentially pave the way for new classes of mechanofunctional active matter systems and collective emergent robotics.

We briefly discuss two swarm robotics examples that exemplify this vision. In the case of small aspect ratios, we have recently demonstrated robophysical analogs of the worm blob.\cite{Ozkan2021} This concept involves individual robots comprising six three-link, two-revolute joints with planar, smart, active particles (smarticles) and two light sensors.\cite{savoie2019robot} This robophysical realization of the blob displays the crawling motion of a biological worm blob by leveraging two major principles: mechanical interactions (entanglements) and function (gait) differentiation. We can consider these synthetic systems as `entangled granular active matter' ensembles.~\cite{savoie2023amorphous}

In a second example, another group utilized large aspect ratio, slender and soft actuators that employ entanglement to hold soft materials.\cite{Beckerstentaclerobot2022} It involved the use of fluidically actuated slender hollow elastomeric filaments that would coil up and entangle around the object being gripped. One can imagine robots similar to these forming entangled blobs and actuating spontaneous tangling and untangling in a similar fashion as the actual worm blob. It would be interesting to investigate the possibility of locomotion and topological transitions in such slender robot blobs.

Achieving a coherent swarm of slender entangled robots for actuation and manipulation of soft objects has only just begun, as new materials and state-of-the-art controller protocols continue to be developed.~\cite{Beckerstentaclerobot2022,savoie2023amorphous,jones2021bubble,savoie2019robot} Guided by the Krogh principle \cite{Krogh1929} stating that ``for such a large number of problems, there will be some animal on which it can be most conveniently studied'', we specifically choose living worms as a model system for studying active polymer physics and the associated emergent collective dynamics in topologically tangled living matter. The worm blob can serve as a guide for designing a decentralized system of soft filamentous robots that can work together. Rather than overlooking topological and steric interactions in robotic design, there is promise in embracing them as a feature for the future of entwined, adaptive, and collective filamentous robotics.\\

\textit{Author Contributions.} A.D. and M.S.B. conceptualized this perspective article. All authors contributed to the writing and revision of the manuscript.\\

\textit{Acknowledgments.} A.D and M.S.B would like to acknowledge all the people involved in fruitful discussions and collaborations: Emily Kaufman, Daniel~Goldman, Yasemin Ozkan-Aydin, Orit~Peleg, Chantal Nguyen, Phil Matthews, and Sara Jabbari-Farouji.
A.~D. warmly thanks Sander Woutersen and Daniel Bonn for their support and insightful discussions. V.~P.~P. acknowledges a Stanford Science Fellowship.
M.~S.~B. acknowledges funding support from NIH Grant R35GM142588 and NSF Grants CAREER IOS-1941933, MCB-1817334, CMMI-2218382 and EF-1935262; the Open Philanthropy Project. A.~D. thanks M.~S.~B for welcoming him in his group and for which he benefits of fruitful ideas and discussions. 

\newpage
\bibliography{Bilbiography.bib}

\end{document}